\documentclass[conference,a4paper,twocolumn]{IEEEtran}

\IEEEoverridecommandlockouts

\ifCLASSINFOpdf
\else
\fi
  \usepackage{tipa}
  \usepackage{graphicx}
	\usepackage{epsfig}
	\usepackage{epstopdf}
	\usepackage{upgreek}
	\usepackage[nolist,nohyperlinks]{acronym}

\usepackage[cmex10]{amsmath}

\usepackage{color, soul}

\usepackage{subfigure}

\usepackage{cite}

\begin{acronym}
\acro{NoC}{Network-on-Chip}
\acro{WNoC}{Wireless Network-on-Chip}
\acro{EM}{Electromagnetic}
\acro{SiO$_2$}{Silicon Dioxide}
\acro{AIN}{Aluminum nitride}
\acro{SiP}{System-in-Package}
\acro{SDM}{Software-defined metamaterial}
\end{acronym}

\begin{document}

% ---------------------------------------------------
% Title
% ---------------------------------------------------


\title{Millimeter-Wave Propagation within a Computer Chip Package}

% ---------------------------------------------------
% Authors, Affiliation, and Acknowledgment
% ---------------------------------------------------

\author{
\IEEEauthorblockN{
Xavier Timoneda\IEEEauthorrefmark{4}, Sergi Abadal\IEEEauthorrefmark{4}, Albert Cabellos-Aparicio\IEEEauthorrefmark{4}, Dionysios Manessis\IEEEauthorrefmark{1},\\
Jin Zhou\IEEEauthorrefmark{3}, Antonio Franques\IEEEauthorrefmark{5}, Josep Torrellas\IEEEauthorrefmark{5}, Eduard Alarc\'{o}n\IEEEauthorrefmark{4} 
}
\IEEEauthorblockA{
\IEEEauthorrefmark{4}\small{NaNoNetworking Center in Catalunya (N3Cat), Universitat Polit\`{e}cnica de Catalunya (UPC), Barcelona, Spain%
}}
\IEEEauthorblockA{
\IEEEauthorrefmark{3}\small{Department of Electrical and Computer Engineering, University of Illinois at Urbana-Champaign (UIUC), Illinois, USA%
}}
\IEEEauthorblockA{
\IEEEauthorrefmark{5}\small{Department of Computer Science, University of Illinois at Urbana-Champaign (UIUC), Illinois, USA%
}}
\IEEEauthorblockA{
\IEEEauthorrefmark{1}\small{System Integration \& Interconnection Technologies, Fraunhofer Institute for Reliability and Microintegration (IZM), Berlin, Germany}\\%
Email: abadal@ac.upc.edu
}
}
%\IEEEauthorblockA{
%\IEEEauthorrefmark{2}Electrical Engineering Department\\
%Iran University of Science and Technology (IUST), Tehran, Iran}


%Telephone: (800) 555--1212, Fax: (888) 555--1212}


\maketitle

% ---------------------------------------------------
% Abstract
% ---------------------------------------------------

\begin{abstract}
Wireless Network-on-Chip (WNoC) appears as a promising alternative to conventional interconnect fabrics for chip-scale communications. The WNoC paradigm has been extensively analyzed from the physical, network and architecture perspectives assuming mmWave band operation. However, there has not been a comprehensive study at this band for realistic chip packages and, thus, the characteristics of such wireless channel remain not fully understood. This work addresses this issue by accurately modeling a flip-chip package and investigating the wave propagation inside it. Through parametric studies, a locally optimal configuration for 60 GHz WNoC is obtained, showing that chip-wide attenuation below 32.6 dB could be achieved with standard processes. Finally, the applicability of the methodology is discussed for higher bands and other integrated environments such as a Software-Defined Metamaterial (SDM).
%Then, two of such novel applications where graphene antennas could play a key role are described.
\end{abstract}

% ---------------------------------------------------
% Keywords
% ---------------------------------------------------

%\begin{IEEEkeywords}
%Wireless Network-on-Chip, 60 GHz radios, flip-chip ceramic BGA package, aperture antenna, patch antenna, monopole antenna, AIN, silicon thickness, channel response, path loss, 1 THz radios, system-in-package.
%\end{IEEEkeywords}

% \fi
\IEEEpeerreviewmaketitle

%Lying between microwave and infrared spectrum, terahertz band is an attractive part of electromagnetic spectrum for space, imaging, spectroscopy and now communication applications. In order to increase the performance of the terahertz systems, efficient transmitter and receiver designs, and terahertz signal processing are key points. Within the framework of the 2017 40th International Conference on Telecommunications and Signal Processing (TSP) held during July 5-7, 2017, Barcelona, Spain, we invite you to contribute by presenting your research and sharing your knowledge on the cutting edge Terahertz Technologies. Topics of interest include, but are not limited to, the following:

%Plasmonic Terahertz sources & receivers
%Photoconductive antennas and photomixers
%Terahertz photonic processing
%IR, THz, and MMW sources and novel generation schemes
%Laser driven terahertz sources
%Quantum cascade lasers (QCLs)
%Resonant tunneling diode (RTD) based terahertz sources and detectors
%Heterojunction bipolar transistors (HBT) based terahertz sources and detectors
%Terahertz instruments, devices and components
%THz and MMW systems, transmission lines and antennas
%Terahertz imaging systems for security applications
%IR, THz, and MMW spectroscopy and material properties
%Ultra high speed MMW digital devices
%MMW and submillimeter-wave radar and communications
%IR, THz and MMW astronomy and environmental science
%IR, THz, and MMW imaging and remote sensing

% ---------------------------------------------------
% Introduction
% ---------------------------------------------------

\acresetall

\section{Introduction} \label{sec:introduction}
Network-on-Chip (NoC) has become the paradigm of choice to interconnect cores and memory within a Chip MultiProcessor (CMP). However, recent years have seen a significant increase in the number of cores per chip and, within this context, it becomes increasingly difficult to meet the communication requirements of CMPs with conventional NoCs alone \cite{Bertozzi2014}. Their limited scalability is in fact turning communication into the next performance bottleneck in parallel processing and, therefore, new solutions are required to avoid slowing down progress in the manycore era \cite{Kim2012Survey}.

Advances in integrated mmWave antennas \cite{Markish2015, Gutierrez2009, Cheema2013} and transceivers \cite{Laha2015, Foulon2014} have led to the proposal of Wireless Network-on-Chip (WNoC) as a potential alternative to conventional NoC fabrics \cite{Matolak2012}. In a WNoC, certain cores are augmented with transceivers and antennas capable of modulating and radiating the information. RF signals propagate through the computing package and can be demodulated by all tuned-in receivers. The main advantage of this approach is that distant cores can communicate with low latency as propagation occurs nearly at the speed of light. In fact, communication is naturally broadcast as long as antennas are roughly omnidirectional. Further, the lack of additional wires between cores provides system-level flexibility not achievable with other interconnects. 

Due to its potential, WNoCs have been investigated extensively from the circuit \cite{Mineo2015, Mondal2017}, link \cite{Mestres2016, Mansoor2015}, network \cite{DiTomaso2015, Abadal2017}, and architecture perspectives \cite{Kim2015, AbadalASPLOS}. However, less attention has been paid to characterizing propagation within the computing package. Modeling the wireless channel is crucial to understand losses, dispersion, and multipath issues that impair communication and impact on the design and performance of the RF transceiver. For instance, the RF amplifiers contribute to more than half of the power consumption in WNoCs \cite{Mondal2017}. Therefore, quantifying channel attenuation becomes essential to optimize the cost of the wireless fabric.

This article investigates the wave propagation inside a realistic flip-chip package by means of EM simulation. We first provide a thorough description of the package and its influence on the antenna placement (Section \ref{sec:modeldescription}). Then, the S-parameters and the path loss over a mesh of 4$\times$4 antennas are obtained at 60 GHz for three different integrated antennas (Section \ref{sec:results}). Through a parametric study, the design of the chip package is optimized to minimize losses and, thus, reduce the cost of wireless communication. Knowledge on the chip package also allows us to propose the space between bumps as a possible propagation channel at higher frequencies (Section \ref{sec:scalability}).

To the best knowledge of the authors, this is the first comprehensive study assuming a realistic model of a flip-chip package and multiple mmWave antennas. Limited experimental measurements in a flip-chip package have been performed below 20 GHz \cite{Guo2002, Branch2005}. At higher frequencies, most works do not discuss the package properly \cite{Zhang2007a, Yan2009, Kimoto2009} or assume wire bonding with the wire acting as antenna \cite{Wu2009}. Little to no work has performed simulation-based explorations \cite{Kimoto2009}.
 
%The remainder of this paper is organized as follows.  presents the system model. Section  evaluates the wireless channel for different antennas and packages, whereas Section \ref{sec:scalability} discusses the scalability and applicability of the methodology. Finally, Section \ref{sec:conclusions} concludes the paper.

%Others have investigated propagation within a computing package \cite{Branch2005, Zhang2007, Kim2016mother}, but the structure differs considerably from that of HSFs.

%In \cite{Yan2009} they investigated the propagation mechanisms in a simplified multilayer chip structure by determining the electric field due to a Hertzian dipole source. They also discussed the effect of a guiding layer placed either below or above the silicon die. In \cite{Kimoto2009} they investigated the power loss between two antennas inside a wire bonding package using on-chip integrated linear dipole antennas placed on the silicon die. They increased the S-$21$ parameter by inserting a dielectric layer below the silicon die and by thinning the silicon thickness.

\section{System Model}
\label{sec:modeldescription}
This work models a flip-chip package with solder bumps for the channel characterization. During its manufacturing process, the solder bumps are deposited on the chip pads, which already carry a valid under bump metallization (UBM) like nickel/gold (Ni/Au). Then, the chip is flipped over and its solder bumps are aligned precisely to the pads of the package carrier external circuit. This is in contrast to wire bonding of chips on the package's carrier (or interposer, we use them interchangeably), in which the chip is mounted upright and wires are used to interconnect the chip pads to external circuitry \cite{Branch2005}. 

Flip-chip packaging is oftentimes preferred over wire bonding for several reasons. First, the I/O signal inductance is much lower due to its much shorter interconnect length, i.e. 100 $\upmu$m or below \cite{Wright2006} compared to 1--5 mm of the wire. The power--ground inductance is also small because the power is brought directly into the core of the silicon die instead of having to be routed to the edges. Further, this approach supports a higher power density since the whole die surface (not just the edges) is used. The edge space originally planned for wire bonding can, in fact, be eliminated to save space and silicon cost.

\begin{figure}[!t]
\centering
\vspace{-0.2cm}
\includegraphics[width=0.9\columnwidth]{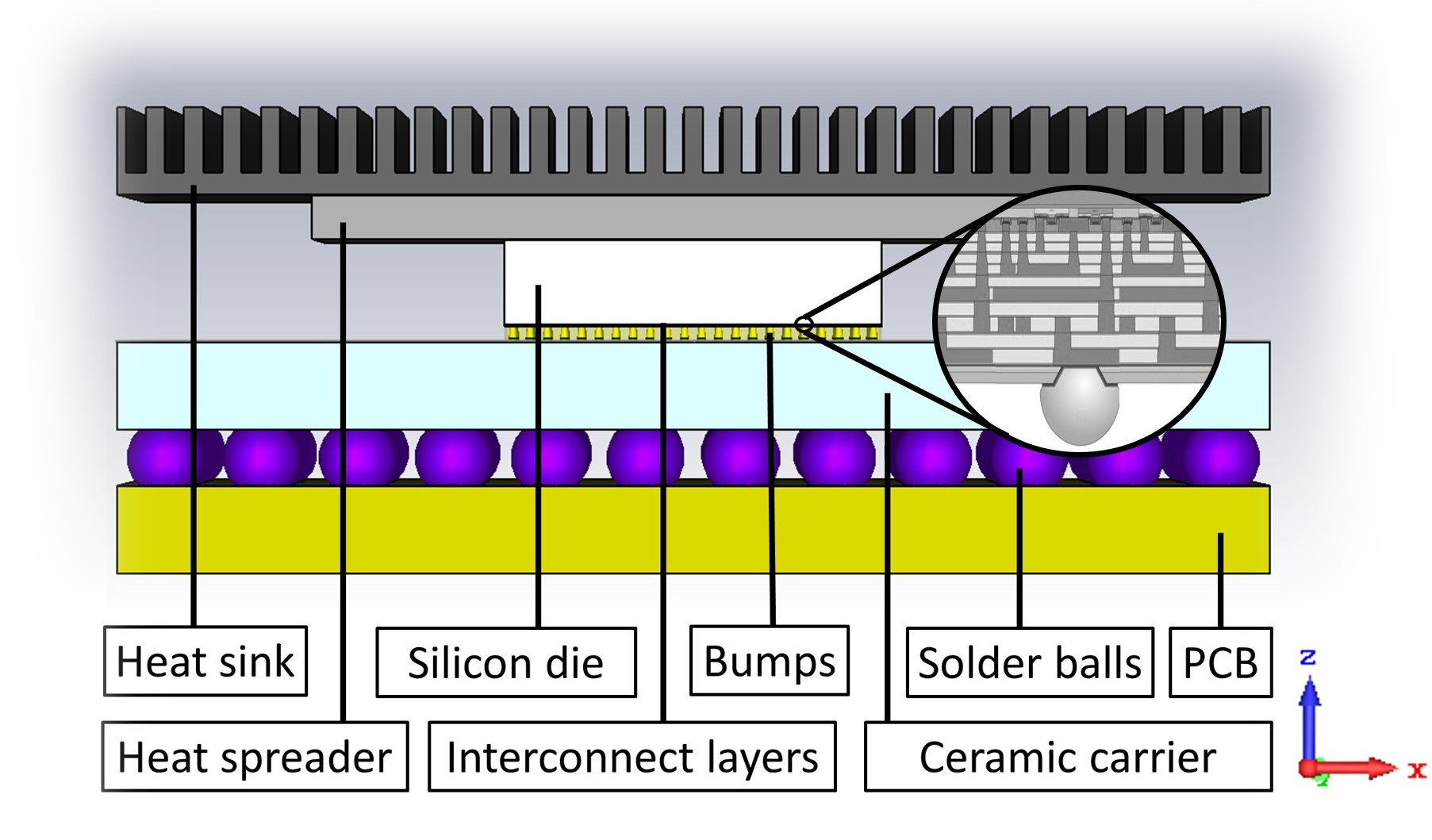}
\vspace{-0.2cm}
\caption{Schematic of the layers of a flip-chip package}
\vspace{-0.4cm}
\label{fig:1}
\end{figure} 

\subsection{Structure of a flip-chip package}
The layers are described from top to bottom following Fig. \ref{fig:1} and Table~\ref{tab:1}. On top, the heat sink and heat spreader dissipate the heat out of the silicon chip, as they both have good thermal conductivity. Bulk silicon serves as the foundation of the transistors. This layer has low resistivity (10 $\Upomega\cdot$cm), which is convenient for the operation of transistors, but not for electromagnetic propagation \cite{Kimoto2009}. The interconnect layers, which occupy the bottom of the silicon die as shown in the inset of Fig. \ref{fig:1}, are generally made of copper and surrounded by an insulator such as silicon dioxide (SiO$_2$) \cite{Markish2015}. 

Very frequently, at the bottom of the interconnect layer, only over the chip I/O pads and being separated by chip passivation, a last UBM is provided (5$\upmu$m Ni/80nm Au) to promote reliable solderability of the solder bumps \cite{Manessis2006}. On the last interconnect layer (13$\upmu$m) and again over the I/O chip pads, the solder bumps are attached and, then, the chip is flipped and soldered to the underlying package carrier, which has on its top a very fine copper metallization external circuitry covered by a solder mask. At the specific pad openings on the ceramic carrier, which are aligned with the solder bumps of the flip chip package, a solderable metallization of also 5$\upmu$m Ni/80nm Au or tin is applied as well. The choice of the carrier material is based on the match of its TCE with that of the silicon die. In this study, the package carrier is a 0.5-mm thick ceramic (other carriers could be epoxy-glass laminate, or silicon). 
%(TCE: ceramic 5-6ppm; epoxy based: 15-18ppm; silicon: 2.5ppm)

At the bottom of the ceramic carrier, an array of other solder balls (320 $\upmu$m) is attached. These are essentially the interconnects of the flip-chip ceramic package with the underlying device PCB substrate, as shown in Fig. \ref{fig:1}. The pitch of the solder balls is around one order of magnitude larger than those of the flip chip bumps. The constellation looks like a 2.5D flip chip on interposer or flip-chip ceramic Ball Grid Array (BGA), prior to its attachment on the device PCB substrate. Although underfilling of the flip-chip bumps or the PCB solder balls is usually performed during manufacturing to increase the reliability of the end-device, such process does not influence our EM model and is considered out of the scope of this paper. %and is therefore considered to be outside the scope of this paper. 

\begin{table}[!t] 
\caption{Characteristics of the layers in a computing package}
\vspace{-0.1cm}
\label{tab:1}
\footnotesize
\centering
\begin{tabular}{lcccc} 
\hline
& {\bf Thickness} & {\bf Material} & {\bf $\varepsilon_{r}$} & tan($\delta$) \\
\hline
Heat sink & 0.5 mm & Aluminum & - & - \\
Heat spreader & 0.25 mm & Thermal cond. & 8.6 & 3$\cdot$10\textsuperscript{-4}\\
Silicon die & 0.489 mm & Bulk Silicon & 11.9 & 0.2517 \\
Interconnections & 13 $\upmu$m & Cu and \textbf{SiO$_2$} & 3.9 & 0.03\\
Bumps & 87.5 $\upmu$m & Cu and Sn & -  & - \\
Ceramic carrier & 0.5 mm & Alumina & 9.4 & 4$\cdot$10\textsuperscript{-4}\\
Solder balls & 0.32 mm & Lead & - & -\\
PCB & 0.5 mm & Epoxy resin & 4 &  - \\
\hline
\end{tabular}
\vspace{-0.5cm}
\end{table}

\subsection{Antenna Placement}
\label{sec:placementant}
The placement of the antenna is discussed for an excitation at 60 GHz. A first option is to place the radiating element as far from the silicon as possible. This is proposed for printed dipoles in several works \cite{Zhang2007a, Zhang2005a, Gutierrez2009}. However, those works do not consider any package and, thus, are not affected by the bumps. In our case, we discard this option because waves would be blocked by the bumps, whose pitch is small compared to the wavelength of the antenna (100 $\upmu$m to 1 mm). 

A second option is to implement the radiating element in the metal layers closest to the silicon. Dipoles \cite{Branch2005} or patch antennas \cite{Yordanov2016} can be places in such layers. In the latter case, which is studied in this paper, the UBM and bumps act as ground plane while the insulator acts as the antenna substrate. 

A third option is to add vertical Through-Silicon Vias (TSV) that act as monopole antennas. Advanced TSV and electroplating techniques \cite{FraunhoferTSV} may allow to adjust the length of the via to make it resonant at the desired frequency. Vertical on-chip monopoles have been proposed recently \cite{Wu2017a}, but using non-standard fabrication and packaging.

\begin{figure*}[!t]
\centering
\includegraphics[width=0.9\textwidth]{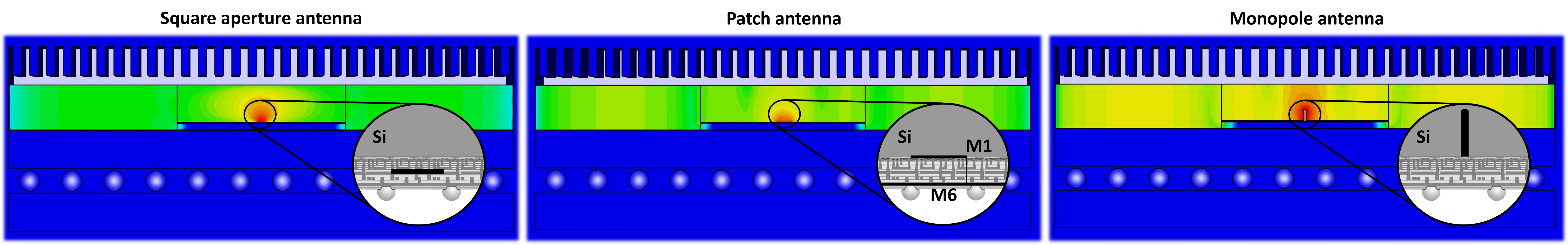}
\vspace{-0.2cm}
\caption{Cross section of the electric field distribution for the three evaluated antennas, whose model is detailed in the insets.}
\vspace{-0.3cm}
\label{fig:2}
\end{figure*}

\subsection{Types of antenna} \label{sec:antennas}

The insets of Fig. \ref{fig:2} show the antennas used in this study. All antennas are tuned to 60 GHz ($|S11|<-10$dB).

\noindent \textbf{Square aperture antenna.} %\label{sec:aperture}
The aperture antenna is modeled as an electrically small waveguide port. It is not a realistic antenna due to the impossibility of building such an ideal aperture, but it serves the purpose of channel characterization as it has a quasi-isotropic radiation diagram.
% due to its dimensions and the fact of not having an interferent structure. 
The antenna is placed horizontally within the SiO$_{2}$.

\noindent \textbf{Patch antenna.} %\label{sec:patch}
The patch antenna is modeled as a planar metallic structure fed from one of the edges. The patch and its ground plane are implemented at the first and last metal layers of the structure. The SiO$_{2}$ serves as the antenna substrate.

\noindent \textbf{Monopole antenna.} %\label{sec:monopole}
The monopole antenna is modeled as a thin and long cylindrical metallic structure, placed vertically passing through the silicon and fed from the first metal layers. This creates a sort of \emph{bed-of-nails} antenna distribution. 

\subsection{Simulation model}
\label{sec:CSTmodel}
The structure shown in Fig. \ref{fig:1} is introduced in a full-wave solver. The dimensions and materials are as listed in Table \ref{tab:1}. The lateral dimensions for the silicon chip and the flip-chip ceramic package are 22 mm and 33 mm, respectively, which are typical values in both research and industry. The interconnect layer stack (13 $\upmu$m) is approximated by a SiO$_2$ layer under the silicon die and a copper layer over the solder bumps, while the antennas are placed either within the SiO$_2$ layer or through the silicon. Such approximation is driven by the rather large metal density close to the bumps and small lateral separation between the interconnects, which at 60 GHz is seen as a solid blocking element. 
%because the separation between the layers of the stack is so small that at 60 GHz their EM properties are equal to a uniform thicker layer of SiO$_2$ and another one of copper.

The full-wave solver allows to obtain the field distribution, the antenna gain, and the coupling between antennas. Then, the channel frequency response $H(f)$ is evaluated as 
\vspace{-0.1cm}\begin{equation} \label{eq:1}
G_{t} G_{r} |H(f)|^{2} = \frac{|S_{21}|^{2}}{(1 - |S_{11}|^{2})\cdot(1 - |S_{22}|^{2})},	
\vspace{-0.1cm}\end{equation}
where $G_{t}$ and $G_{r}$ are the transmitter and receiver antenna gain, $S_{21}$ is the coupling between transmitter and receiver, whereas $S_{11}$ and $S_{22}$ are the reflection coefficients at both ends \cite{Lin2007}. Once evaluated, a path loss analysis can be performed by fitting the attenuation over distance to
\vspace{-0.1cm}\begin{equation} \label{eq:2}
L_{dB} = 10n \cdot \log_{10}(d) + C,
\vspace{-0.1cm}\end{equation}
where $d$ is the distance between antennas and $n$ is the path loss exponent \cite{Zhang2007a}. The path loss exponent is around 2 in free space, below 2 in guided or enclosed structures, and above 2 in lossy environments. 

\section{Simulation results}
\label{sec:results}
We use CST \cite{CST} to obtain the field distribution and S-parameters in the 55--65 GHz band for a homogeneous distribution of 4$\times$4 antennas within the package.

\begin{figure}[!t]
\centering
\vspace{-0.2cm}
\includegraphics[width=0.8\columnwidth]{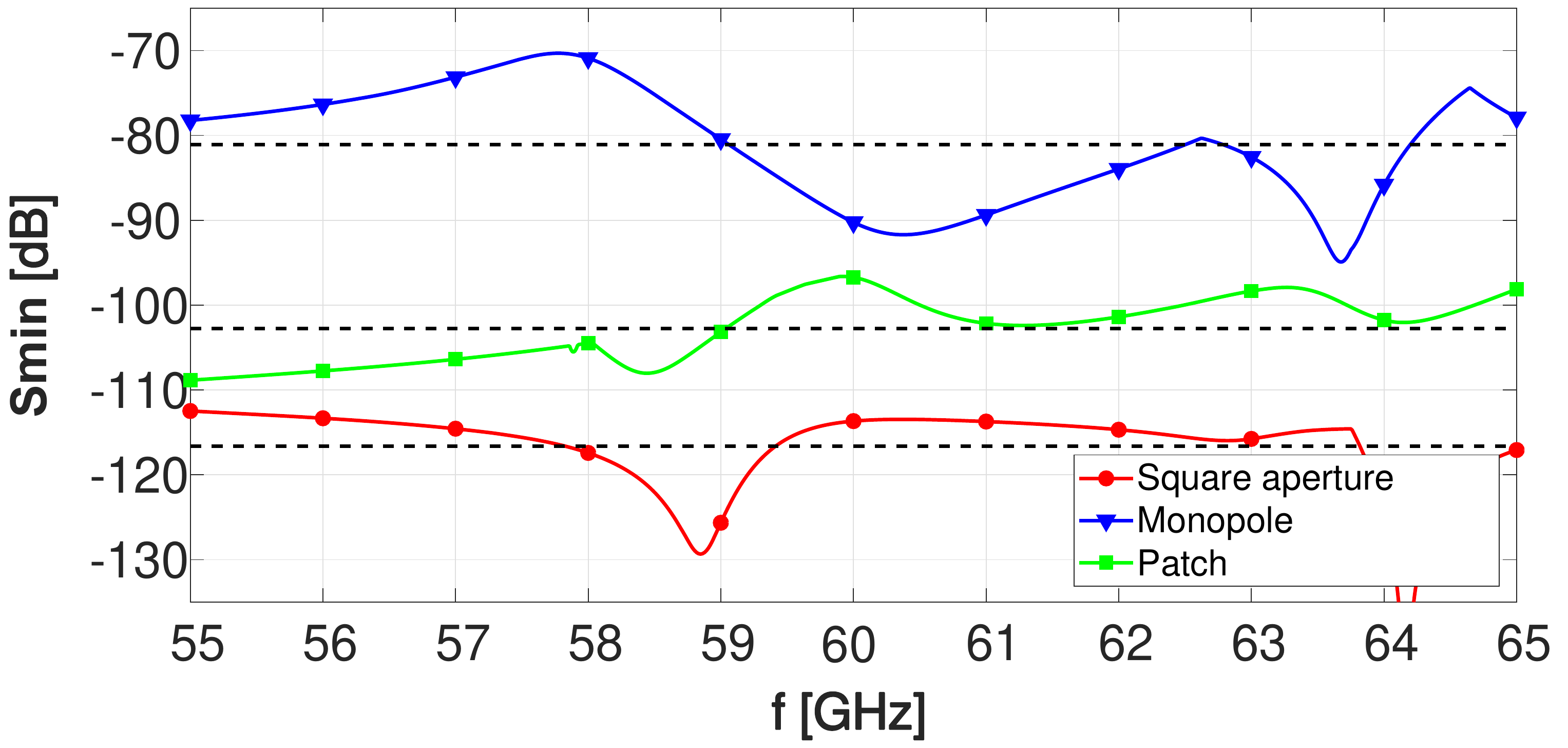}
\vspace{-0.4cm}
\caption{Worst-case S-Parameter for each antenna. Dashed lines represent the mean over the 55--65 GHz band.}
\label{fig:3}
\vspace{-0.4cm}
\end{figure} 

\subsection{Antenna comparison} \label{sec:antenna}
The field distribution for each type of antenna is shown in Fig. \ref{fig:2}. It is observed that the field is confined at the region between the heat sink and the ceramic carrier. This is caused by the presence of several metallic blocks: heat sink on top, UBM at the bottom, and package walls on the side. The aperture and patch, being planar, radiate mostly towards the heat sink and signals propagate laterally due to multiple reflections. On the other hand, the monopole has an azimuthal radiation pattern --most of the power propagates laterally. 

Fig. \ref{fig:3} plots the worst-case coupling $S_{min}$, which is the minimum S-parameter between any antenna pair 
\vspace{-0.1cm}\begin{equation}
S_{min}(f) = \min_{i,j\neq i}{S_{ij}(f)},
\vspace{-0.1cm}\end{equation}
as well as the average value over the whole band. The results show that, in consonance with the field distribution analysis, the monopole has the best coupling among the considered alternatives (-81.1 dB). The coupling between patches is, as expected, higher than for the electrically small --and therefore inefficient-- aperture antenna. In all cases, the very large attenuation is mostly due to the presence of lossy silicon.

The standard deviation of $S_{min}$ over frequency is also evaluated. A low value is preferred as it implies a larger effective bandwidth. High values are due to notches produced by either the resonant nature of the antennas or multipath effects. The monopole and the patch yield the highest (6.03 dB) and lowest value (3.66 dB), suggesting the existence of an interesting efficiency--bandwidth tradeoff among antennas.

%The pros and cons of the different antennas were studied analyzing the minimum S-Parameter for each case, shown in Figure . The trade-offs are summarized in Table \ref{tab:2}. The monopole antenna has the highest mean but has the highest variance. This is due to the non-optimal adaptation of the antenna. The mean of the monopole, even that is higher than with the other kinds of the antenna, is still very low (-81.1 dB) due to antenna inefficiency when it is surrounded by a big piece of silicon, and needs to be raised to have acceptable path loss results. The techniques to increase the S-Parameters are chosen, explained, tested and optimized in the following sections.

%\begin{table}[!b] 
%
%\caption{Trade-off for each type of antenna}
%\vspace{-0.1cm}
%\label{tab:2}
%\footnotesize
%\centering
%\begin{tabular}{lccc} 
%\hline
 %& {\bf Pros} & {\bf Cons} \\
%\hline
%Square aperture antenna &  & High variance, lowest mean\\
%Monopole antenna & Highest mean & Highest variance\\
%Patch antenna & Lowest variance & Lower mean than monopole\\
%
%\hline
%\end{tabular}
%\vspace{-0.5cm}
%\end{table}

\subsection{Package design} \label{sec:design}
The attenuation values obtained above need to be greatly improved to consider intra-chip wireless as a viable and efficient option. Package co-design techniques that may help to address this issue are explained and evaluated next.

\noindent \textbf{Additional dielectric layer.} %\label{sec:design}
By considering the field distribution results above and basing on other studies \cite{Guo2002, Yan2009}, a good way to reduce propagation losses is to make use of the heat spreader as it generally has low electrical losses \cite{Kimoto2009}. Silicon carbide, beryllium oxide and aluminum nitride (AIN) are widespread due to their excellent thermal properties; in this study, AIN is chosen as it has the lowest electrical losses. 

\begin{figure}[!t]
\centering
\subfigure[\label{fig:4a}Improvement over no AIN]{\includegraphics[width=0.45\columnwidth]{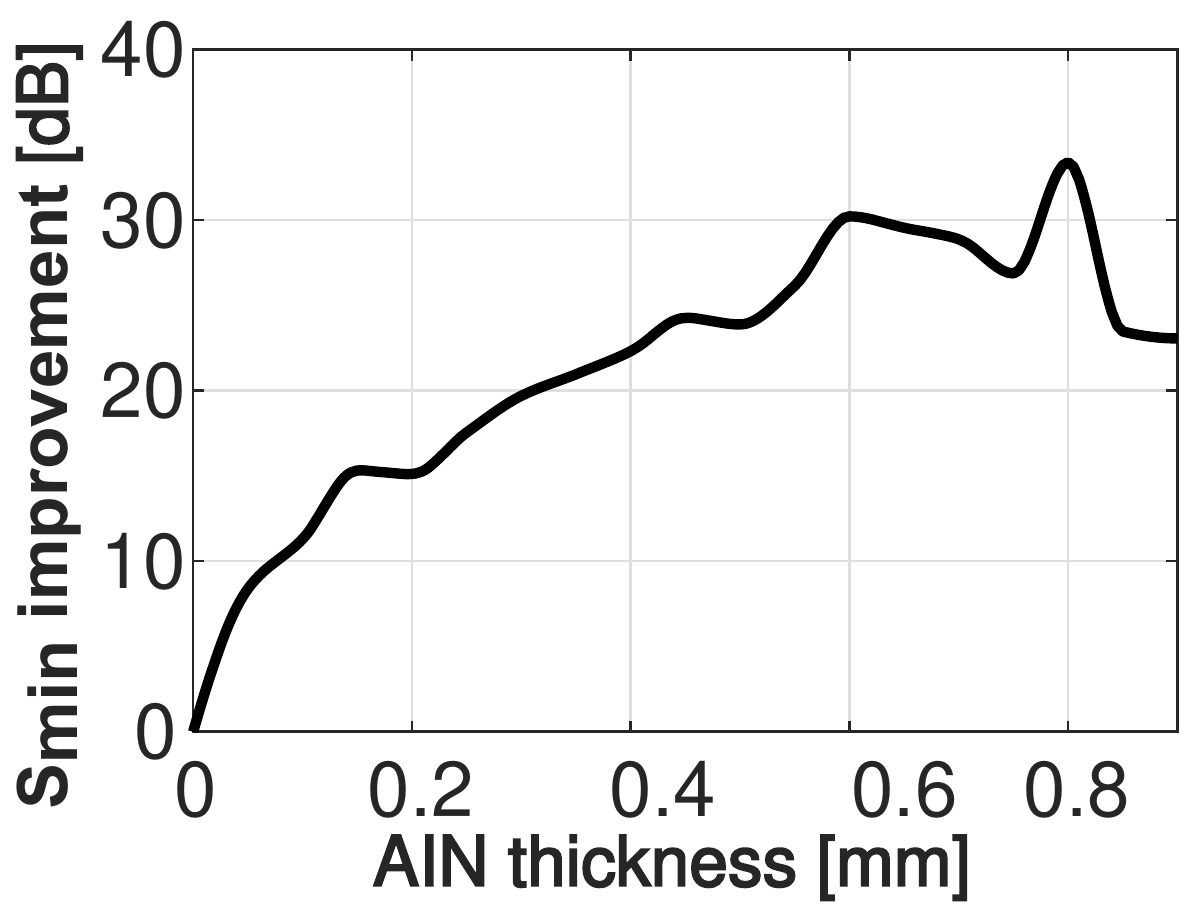}}
\subfigure[\label{fig:4b}Improvement over 0.7-mm Si]{\includegraphics[width=0.45\columnwidth]{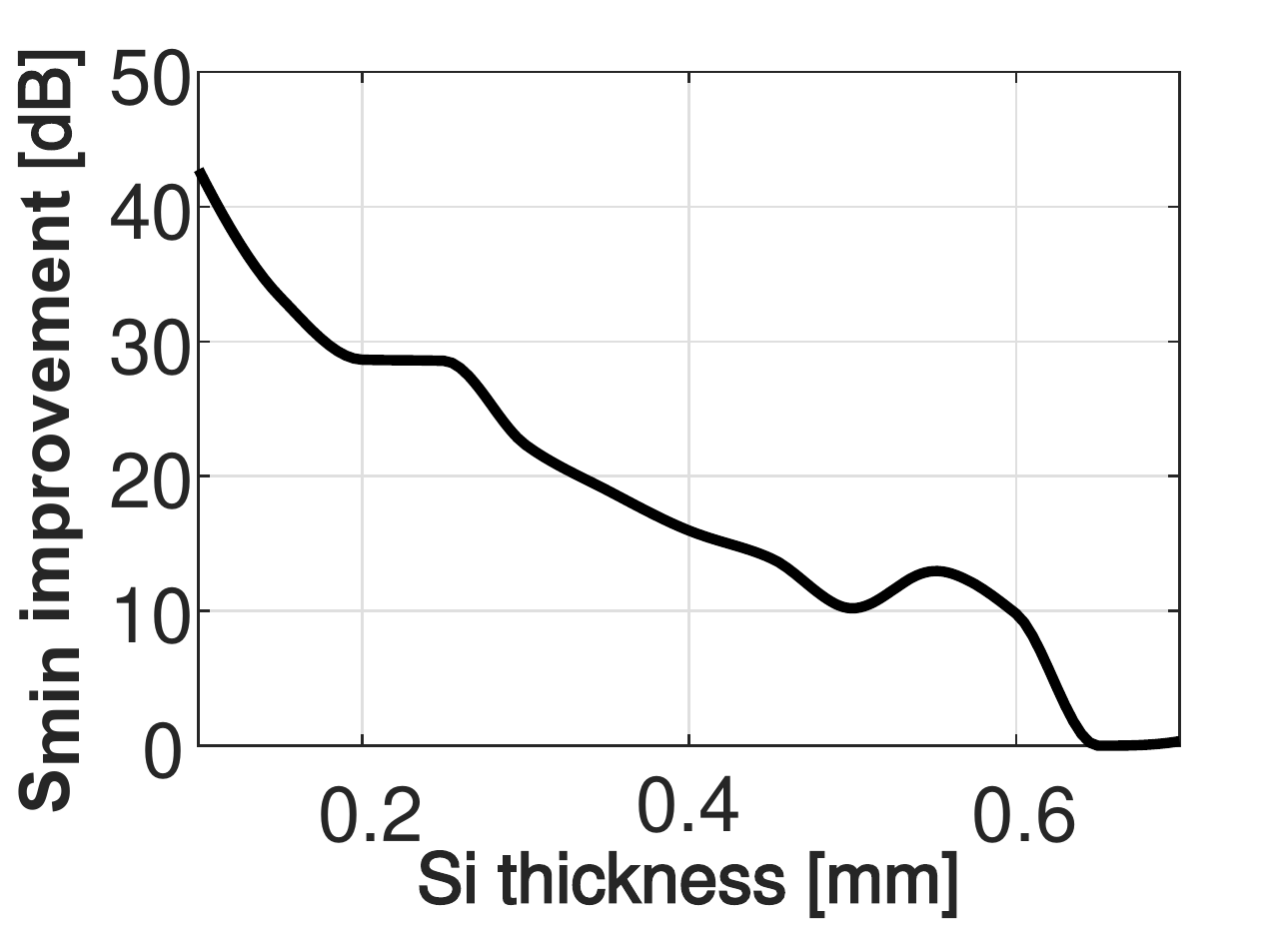}}
\vspace{-0.3cm}
\caption{Average coupling enhancement for different AIN and Si thicknesses.}
\vspace{-0.3cm}
\end{figure}

Simulations are repeated for different AIN thicknesses. As observed in Fig. \ref{fig:4a}, increasing the AIN thickness improves the average $S_{min}$ up to 33 dB with respect to not having AIN. Although not shown here, it is worth noting that the standard deviation of $S_{min}$ oscillates (not uniformly) between 2.7 and 8.2 dB. Therefore, it is a parameter to take into account when selecting the AIN thickness.

\noindent \textbf{Thinning the silicon.} %\label{sec:design}
The characteristics of the bulk silicon suggest that most losses occur in it. Consequently, reducing the silicon thickness is an intuitive way to minimize the losses as it shortens the propagation through the silicon layer. Furthermore, the radiation efficiency of the antennas increases as the near-field influence of the lossy silicon die is minimized. 

To quantify these effects, simulations are repeated by considering silicon thicknesses down to 100 $\upmu$m, although chip makers can reportedly reduce that further to tens of microns \cite{Bieck2010}. As we can see in Fig. \ref{fig:4b}, the path loss difference between the 0.1mm and 0.7mm cases is over 40 dB. The standard deviation oscillates between 2.9 and 6.6 dB and shall be considered when designing the package.

\noindent \textbf{Antenna and package co-design.} %\label{sec:design}
Previous results have provided a choice for the antenna, as well as rough dimensions for the silicon and heat spreader layers that minimize the path loss. To provide a design point that adds up the benefits of the three processes, we perform an antenna-package co-design optimization. We explore the design space around the optimal silicon and AIN thicknesses by keeping the monopole matched at 60 GHz at all times. This is important because the effective wavelength of the antenna slightly varies among cases. 

\begin{figure}[!t]
\centering
\includegraphics[width=0.9\columnwidth]{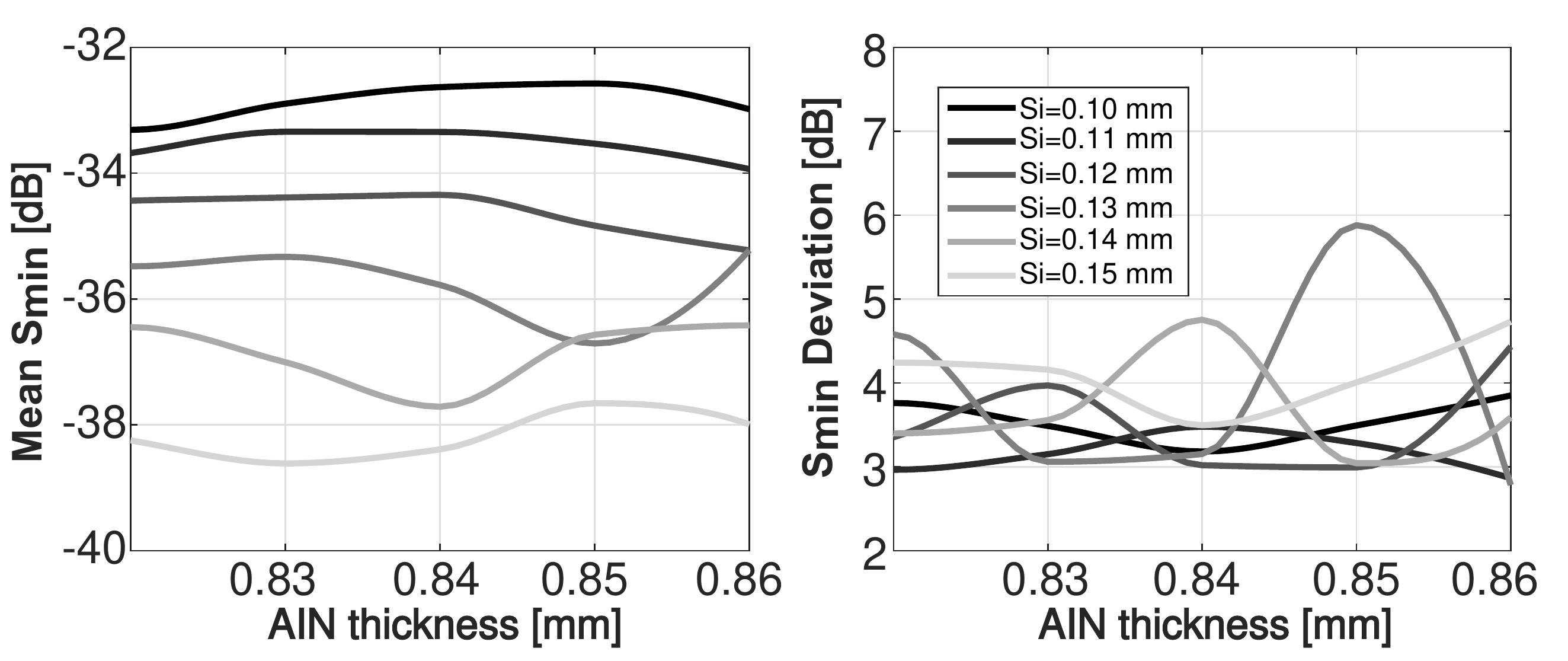}
\vspace{-0.35cm}
\caption{Mean and standard deviation of the worst-case coupling for several silicon and AIN thicknesses.}
\vspace{-0.3cm}
\label{fig:5}
\end{figure}

Fig. \ref{fig:5} shows the results of the parametric design-oriented study. It is found that the optimal silicon and AIN thicknesses are 0.10 mm and 0.85 mm, respectively, as they yield the highest mean with low variance. In summary, the optimization process has reduced the losses from 81.1 dB to only 32.6 dB.

%\begin{figure}[!t]
    %\centering
    %\begin{subfigure}[aaa]{\textwidth}
        %\includegraphics[width=0.23\textwidth]{./Figures/Figure4a.eps}
    %\end{subfigure}
		%\begin{subfigure}[vvv]{\textwidth}
        %\includegraphics[width=0.23\textwidth]{./Figures/Figure4b.eps}
    %\end{subfigure}
		%\caption{(a) Improvement over zero AIN thickness and (b) Improvement over 0.7mm silicon.}
		%\label{fig:4}
%\end{figure} 

%\vspace{-0.2cm}
\subsection{Path loss analysis} \label{sec:together}%\vspace{-0.2cm}
\begin{figure}[!t]
\centering
\includegraphics[width=0.75\columnwidth]{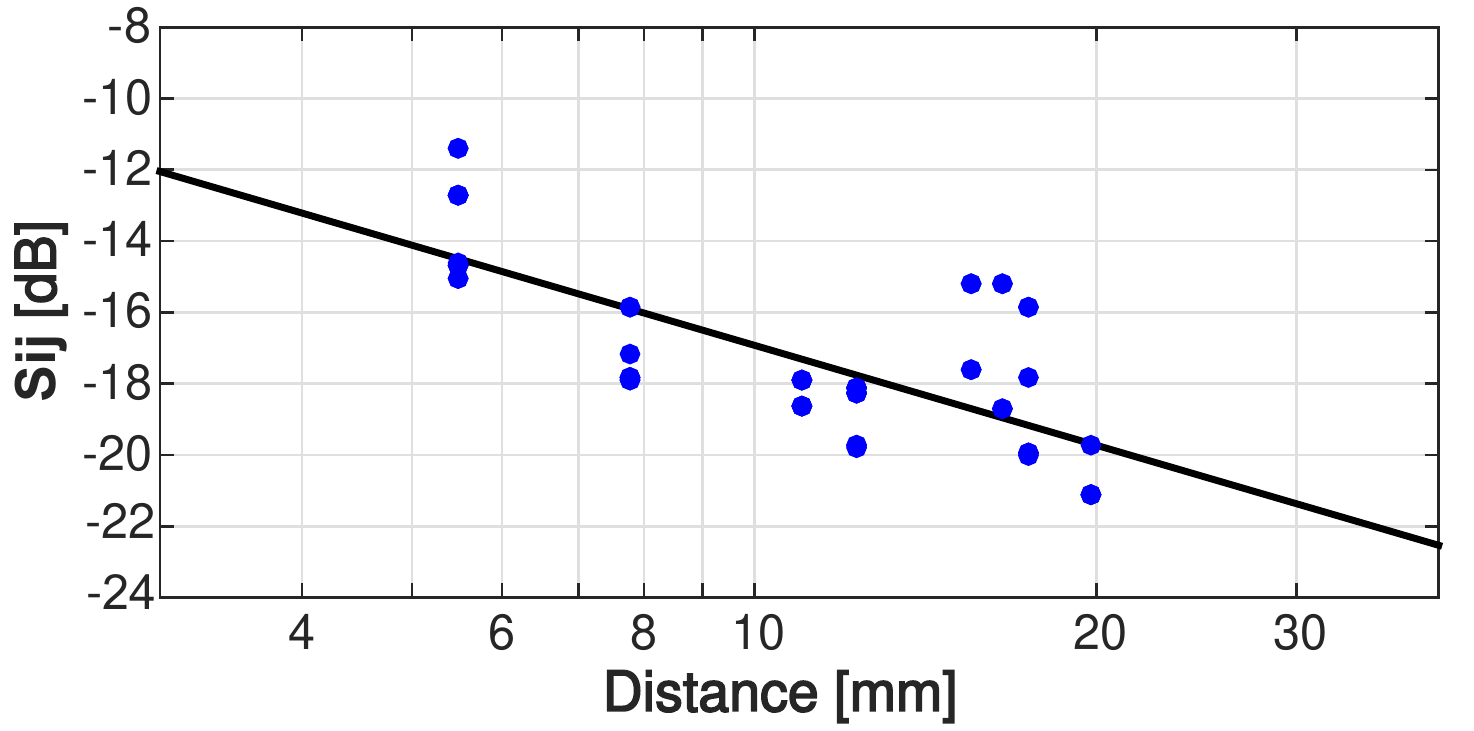}
\vspace{-0.35cm}
\caption{Path loss for each antenna pair and log-distance fitting.}
\label{fig:6}
\vspace{-0.3cm}
\end{figure}

We next evaluate the path loss exponent for the optimal case found with the antenna-package co-design. The channel response is computed for every antenna pair using Eq. \eqref{eq:1} assuming identical gains in transmission and reception. The attenuation at 60 GHz is plotted as a function of the distance between antennas and a linear regression fitting is performed with distance in a logarithmic scale. Fig. \ref{fig:6} shows the results: a line with a slope of 9.32 dB/decade is obtained, which means that the path loss exponent is $n=0.932$. This is significantly lower than the freespace exponent ($n=2$) and the exponent obtained in \cite{Zhang2007a} for on-chip mmWave propagation without a chip package ($n\approx 1.4$). This result stresses the importance of the enclosed nature of the package and its waveguiding effect. %This will probably have an impact on the multipath, which we leave for future work.

%\vspace{-0.2cm}
\section{Discussion}%\vspace{-0.2cm}
\label{sec:scalability}
\noindent \textbf{Other scenarios.} The methodology of this work is applicable to new relevant scenarios such as System-in-Package (SiP) or Software-Defined Metamaterials (SDM). In a SiP \cite{Shamim2017}, several flip chips can be assembled on the ceramic carrier, making the package more of an integration platform for heterogeneous functionalities (e.g., CPU+GPU). In a SDM \cite{AbadalACCESS, Liaskos2015, Liu2018ISCAS}, a network of controllers are co-integrated below a metasurface to provide EM programmability. Both scenarios shall incorporate the effect of the added components into the EM model.

\begin{figure}[!t]
\centering
\includegraphics[width=0.85\columnwidth]{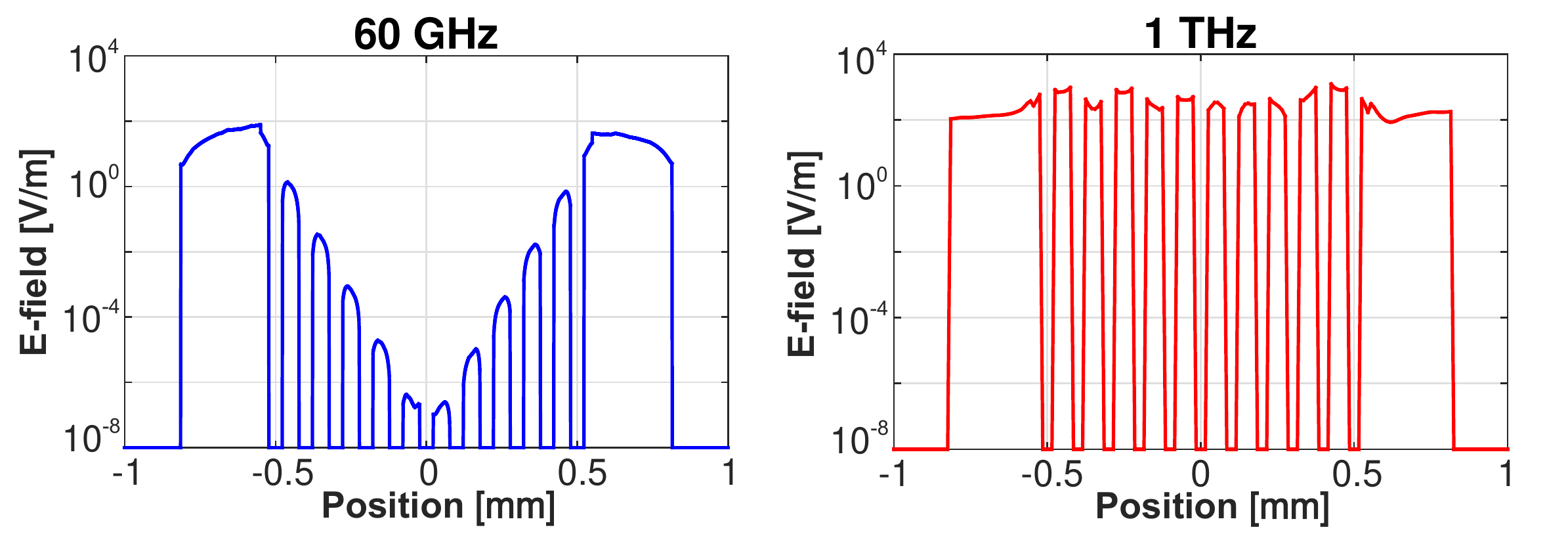}
\vspace{-0.4cm}
\caption{Field distribution along the X-axis within the bump region.}
\vspace{-0.5cm}
\label{fig:7}
\end{figure}

\noindent \textbf{Scaling in frequency.} The location of the antennas in this work has been motivated by the small pitch of the bumps. The blocking effect of the bumps is observed in the left plot of Fig. \ref{fig:7} as the electric field along the X-axis decays several orders of magnitude in the space between bumps (it is null within them). As the wavelength becomes commensurate to the pitch, however, one would expect bumps to no longer be an obstacle. This is confirmed in the right plot of Fig. \ref{fig:7}, which assumes an excitation at 1 THz ($\sim$100 $\upmu$m) and shows strong fields in the center of the die. This suggests that as CMOS or graphene-based THz technologies become available \cite{Kim2016mother, Khamaisi2016, AbadalMICRO}, propagation through the bumps region should be considered.

%\vspace{-0.2cm}
\section{Conclusions}%\vspace{-0.2cm}
\label{sec:conclusions}
In this work, we performed a simulation-based study of wave propagation within a realistic flip-chip ceramic BGA package. Among the evaluated antennas, the vertical monopole delivered higher coupling. It has been also demonstrated that by thinning the lossy silicon and using the heat spreader as propagation layer, losses between monopole antennas can be reduced by $\sim$50 dB. The path loss analysis yielded an exponent of 0.9, which confirms that the waveguide effect is dominant in this environment. Finally, it has been suggested that inter-bump propagation could be feasible at THz frequencies and should be considered in emerging applications.

%\vspace{-0.1cm}
\section*{Acknowledgment}
%\vspace{-0.1cm}
This work was supported by the Spanish MINECO (PCIN-2015-012), the EU's H2020 FET-OPEN program (grant 736876), and by NSF (CCF 16-29431).


% ---------------------------------------------------
% References
% ---------------------------------------------------

\bibliographystyle{IEEEtran}
\bibliography{IEEEabrv,references}
%

% ---------------------------------------------------
% End of Document
% ---------------------------------------------------

\end{document}